%% file: Preprint_LQ2pairs_DZero_v2-0.tex
\documentclass[prd,twocolumn,showpacs,groupedaddress]{revtex4}

\usepackage{graphicx}
\usepackage{dcolumn}
\usepackage{bm}

\newlength{\picwidth}
\newlength{\picoffset}
\newcommand{\SP}{\!}
\newcommand{\VSP}{\vspace{0.1cm}}
\newcommand{\VSPP}{\vspace{0.1cm}}
\newcommand{\gev}{\,\rm GeV}

\newcommand{\invpb}{\,\rm{pb}^{-1}}

\newcommand{\met}{$E\!\!\!\!/_{T}$}
\newcommand{\pythia}{{\sc pythia}} 
\newcommand{\alpgen}{{\sc alpgen}} 
\newcommand{\geant}{{\sc geant}} 

\newcommand{\LQ}{LQ}
\newcommand{\Bf}{\mathcal{B}} 
\newcommand{\R}{\mathcal{R}} 

\newcommand{\MyDate}{January 31, 2005}

\newcommand{\totallumiB}{\mbox{$294\pm 19\invpb$}}

\newcommand{\masslimit}{\mbox{$m_{\LQ_2} > 247\gev$}}
\newcommand{\masslimitbetahalf}{\mbox{$m_{\LQ_2} > 182\gev$}}
\newcommand{\combmasslimit}{\mbox{$m_{\LQ_2} > 251\gev$}}
\newcommand{\combmasslimitbetahalf}{\mbox{$m_{\LQ_2} > 204\gev$}}
\newcommand{\expectedmasslimit}{\mbox{$m_{\LQ_2}^{\text{expected}} > 251\gev$}}
\newcommand{\expectedmasslimithalf}{\mbox{$m_{\LQ_2}^{\text{expected}} > 199\gev$}}

\begin{document}



\title{
  Search for Pair Production of Second Generation Scalar 
  Leptoquarks \\in {\boldmath $p\bar{p}$} Collisions at 
  {\boldmath $\sqrt{s}=1.96\,\rm{TeV}$}
}


\input list_of_authors_r2.tex
%
\date{\MyDate}

\begin{abstract}
  \vspace*{-0.3cm} 
    We report on a 
    search for the pair production of
    second generation scalar leptoquarks ($\LQ_2$) 
    in $p\bar{p}$ collisions  
    at the center-of-mass energy $\sqrt{s} = 1.96$ TeV, using 
    data corresponding to 
    an integrated luminosity of \totallumiB{} recorded with the 
    D\O\ detector.
    No evidence for a leptoquark signal in the 
    $\LQ_2\overline{\LQ}_2\rightarrow\mu q\mu q$ channel 
    has been observed, and  
    upper bounds on the product of cross section times branching
    fraction were set.
    This yields lower mass limits of \masslimit{} for 
    $\beta=\Bf(\LQ_2\rightarrow \mu q)=1$ and 
    \masslimitbetahalf{} for $\beta=1/2$. 
    Combining these limits with previous D\O\ results, 
    the lower limits on the mass of a 
    second generation scalar leptoquark are \combmasslimit{} and 
    \combmasslimitbetahalf{} for $\beta=1$ and $\beta=1/2$, respectively.
    \vspace{-0.5cm}

\end{abstract}



\hspace{5.2in}\mbox{FERMILAB-PUB-06/017-E}

\maketitle

\newpage

Leptoquarks, colored bosons which carry both lepton ($l$) and quark ($q$)
quantum numbers and third-integer electric charge, appear in several
extensions of the standard model of particle physics \cite{theory}.
Leptoquarks 
could, in principle, decay into any combination of a lepton and a quark. 
Experimental limits on lepton number violation, 
on flavor-changing neutral currents, and on proton decay, however, 
motivate the 
assumption that there would be three different generations of leptoquarks. 
Each of these leptoquark generations couples to only one generation of
quarks and leptons, and, therefore, conserves the corresponding lepton
and quark family numbers \cite{theory2}. 
As a consequence, leptoquark masses could be as low as 
$\mathcal{O}(100\gev)$, allowing the production of leptoquarks in reach of
present collider experiments.

At the Tevatron collider, leptoquarks would be produced in pairs,
primarily through $q\bar{q}$ annihilation and gluon fusion. These
production mechanisms would be independent of the unknown coupling
$\lambda$ between the leptoquark, the lepton, and the quark.

This analysis focuses on the search for pair-produced second
generation scalar leptoquarks ($\LQ_2$) in $p\bar{p}$ collisions at 
$\sqrt{s}=1.96\,\rm{TeV}$. 
Assuming 100\% branching fraction to a charged lepton and a quark, $\beta =
\Bf(\LQ_2\rightarrow\mu q) = 1$, a pair of second generation 
leptoquarks, $\LQ_2\overline{\LQ}_2$,
decays into two
muons and two 
quarks.
This decay will have no missing transverse energy. 
For $\beta=1/2$, the same final state is produced $25\%$ of the time.
The D\O\ collaboration published $95\%$ confidence level (C.L.) 
mass limits for 
second generation scalar leptoquarks of $m_{\LQ_2}>200\gev$ 
($180\gev$) for $\beta=1$ ($1/2)$ at 
$\sqrt{s} = 1.8\,\rm{TeV}$, using $94\invpb$ of Run I Tevatron data 
\cite{run1_lq2pairs}.
Recent CDF analyses of dimuon + jet and single muon + jet Run II Tevatron
data give $m_{\LQ_2}>226\gev$ ($208\gev$) for $\beta = 1$ ($1/2$), 
determined from $198\invpb$ of data \cite{run2_cdf}. 

The D\O\ Run II detector \cite{run2det} is composed of several 
layered elements. Nearest the beam is a
central tracking system consisting of a silicon
microstrip tracker (SMT) and a central fiber tracker (CFT), both
located within a $2\,\rm{T}$ superconducting solenoidal
magnet. Muon momenta are measured from the
curvature of muon tracks in the central tracking system. 
Jets are reconstructed from energy depositions in the 
three liquid-argon/uranium
calorimeters 
outside the tracking system: 
a central section (CC) covering up to $|\eta|\approx
1.1$ and two end calorimeters (EC) extending coverage to $|\eta|\approx
4$, all housed in separate cryostats, 
where
$\eta=-\ln{\left(\tan{\frac{\theta}{2}}\right)}$ denotes the pseudorapidity and
$\theta$ is the polar angle with respect to the proton beam direction. 
Scintillators located between the CC and EC cryostats
provide sampling of hadron showers for $1.1<|\eta|<1.4$. 
A muon system beyond the 
calorimeters
consists of a layer
of drift-tube tracking detectors and scintillation trigger 
counters before $1.8\,\rm{T}$ iron
toroids, followed by two additional similar layers after the toroids 
\cite{muonsystem}.

The data used in this analysis were collected 
during Run II of the Fermilab Tevatron collider
between August 2002 and July 2004 
and correspond to an  
integrated luminosity of \totallumiB{}. 
The sample of candidate events used in this search was collected
with a set of triggers that required either one or two muon candidates
in the muon system.
The trigger efficiency for the $\mu j\mu j$ events considered in this
analysis was measured to be $(89\pm3)\%$.

Muons in the region $|\eta|<1.9$ were reconstructed offline from hits 
in the three layers of the muon
system which were matched to isolated tracks in the
central tracking system 
to remove the background from
heavy-quark production.
This muon isolation was assured by requiring
the sum of the transverse momenta of all other tracks in a
$\Delta\R=\sqrt{(\Delta\phi)^2+(\Delta\eta)^2}<0.5$ cone 
around the muon to be smaller than $4\gev$, where $\phi$ is the
azimuthal angle around the direction of the incident beam.
Cosmic ray muons were rejected by cuts on
the timing in the muon scintillators and by removing back-to-back
muons. Jets were reconstructed using the iterative, midpoint cone 
algorithm \cite{conealgo} with a cone size of $\Delta\R=0.5$. 
The jet energies
were calibrated as a function of the jet transverse energy and $\eta$
by balancing the transverse energy in photon plus jet events.
Requiring $|\eta|<2.4$ for all jets removes the QCD background
from events with jets at very small angles to the beam direction and,
therefore, with large cross sections.

\begin{figure*}
\includegraphics[width=1.9\picwidth]{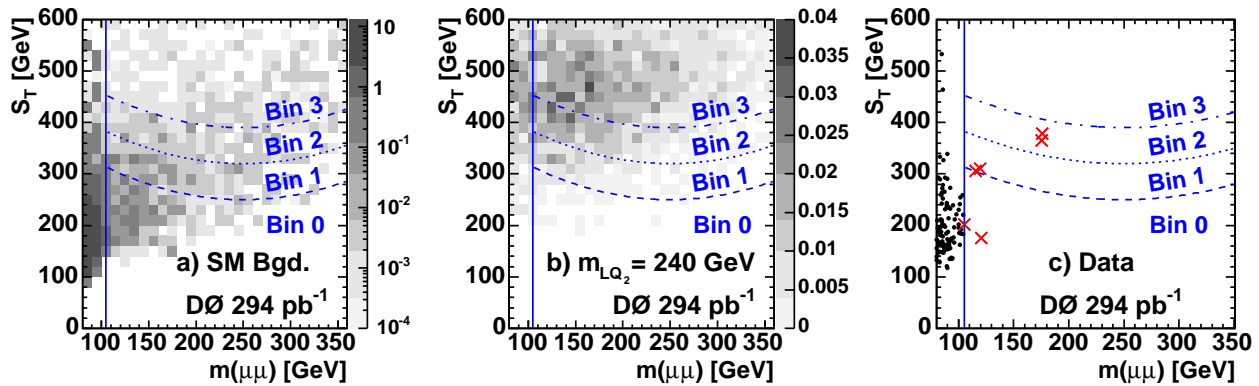}
\vspace*{-0.3cm} 
\caption{\label{fig:STvsM} 
  Scalar sum of the transverse energies, $S_T$, 
  as a function of the dimuon
  mass: a) for the SM background, b) for 
  leptoquark signal with mass 
  $m_{\LQ_2}=240\gev$ and $\beta=1$, and c) for data (the six 
  events surviving the $Z$ boson veto are highlighted).
  The vertical line illustrates the $Z$ boson veto and the curved lines
  show the boundaries between the signal bins (see text for definition).
  The distributions shown in a) and b) are normalized to the 
  integrated luminosity.
}
\end{figure*}

The background is dominated by 
the Drell-Yan (DY) events in the channel $Z/\gamma^*\rightarrow\mu\mu$ 
(+jets).
QCD multijet events faking muons are suppressed
by  the isolation requirement and the thick shielding of the muon
detectors.
To evaluate the
contribution from DY background, samples of Monte Carlo (MC) events
were generated with \pythia{} \cite{pythia}. 
The number of \pythia{} events was normalized to yield the predicted 
next-to-next-to-leading order (NNLO) 
cross section \cite{xsection} at the $Z$-boson resonance. 
The events were furthermore reweighted as a function of the dimuon mass 
in order to describe the NNLO prediction for the differential
cross section $d\sigma/dm_{\mu\mu}$ \cite{xsection}.
An additional sample, generated with \alpgen{} \cite{alpgen} and 
based on a matrix-element calculation for $Zjj$, was used to test
systematic uncertainties due to the shape of the jet transverse 
energy distribution.
Samples of \pythia{} $t\bar{t}$ ($m_t=175\gev$) and
$WW$ samples were used to estimate the background
contributions from top quark and $W$ boson pair production.
The signal efficiencies were calculated using 
samples of $\LQ_2\overline{\LQ}_2\rightarrow\mu q\mu q$ events
simulated with \pythia{} 
for leptoquark masses from $140\gev$ to $300\gev$ in steps of $20\gev$.
All Monte Carlo events 
were generated using CTEQ5L \cite{cteq5} parton distribution 
functions (PDFs)
and processed using a full simulation of the
D\O\ detector based on \geant{} \cite{geant} and the D\O\ event
reconstruction \cite{run2det}.

Offline, events were required to have two muons with 
transverse momenta $p_T$ exceeding
$15\gev$ and at least two jets with transverse energies $E_T$ greater than
$25\gev$. 
The momentum resolution degrades with increasing $p_T$, and hence the 
resolution on the dimuon mass $m(\mu\mu)$ with increasing
$m(\mu\mu)$. Therefore, in order to reduce the DY background at high 
$m(\mu\mu)$ 
and to account for muon tracks with large momentum uncertainty, 
corrections were applied to the muon 
momenta by taking advantage of the fact that no missing transverse energy
is expected in either signal or DY events. 
The missing transverse energy \met{} was 
estimated from the transverse energy balance of all
muons and jets ($E_T>20\gev$) in the event.
The momentum of the muon most opposite to the 
\met{} direction in the $r$-$\phi$ plane (i.e.\ in the plane perpendicular 
to the incident beam) was rescaled such that the component of the 
missing transverse energy parallel to the muon vanished. 
This correction suppressed the contribution from $Z$ boson events 
misreconstructed in the high mass region where the search for
leptoquarks took place. 
To further reduce the background from DY
events a $Z$ boson veto cut (dimuon mass $m(\mu\mu)>105\gev$) was applied.
Six events survive this last cut, 
while $6.8\pm2.0$  
events are expected
from standard model backgrounds, which mainly consists of 
DY ($6.1\pm2.0$) and  $t\bar{t}$ ($0.69\pm0.07$).

\begin{figure}
\includegraphics[width=\picwidth]{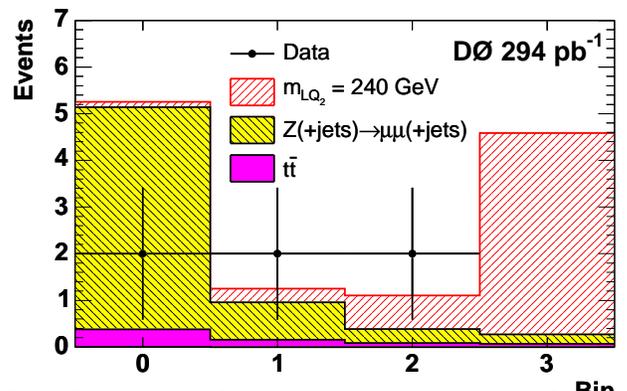}
\vspace*{-0.7cm} 
\caption{\label{fig:lastcut} 
  Distribution of events over the four bins as defined in 
  the text for a scalar leptoquark with mass 
  $m_{\LQ_2}=240\gev$ and $\beta=1$.
}
\end{figure}

The remaining events after the $Z$ boson veto cut
were arranged in four bins.  
Second generation leptoquark
events are expected to
have both high dimuon masses and large values of $S_T$, which is the
scalar sum of the transverse energies of the two highest-$p_T$ muons
and the two highest-$E_T$ jets in the event, as can be seen in 
Fig.\ \ref{fig:STvsM}b) for a leptoquark mass of $240\gev$.
The separation between bin $i$ and bin $i-1$, 
$i \in \left\{1,2,3\right\}$, is defined as:
\begin{displaymath}
  S_T > \frac{0.003}{\rm{GeV}}\cdot(m(\mu\mu)-250\,\rm{GeV})^2+180\,\rm{GeV} + \text{$i$} \cdot70\,\rm{GeV}.
\end{displaymath}
This binning, 
which effectively results in bins in the order of increasing $S/B$,
is illustrated by the curved lines in Fig.\ \ref{fig:STvsM} 
for the expected standard model backgrounds, an example $\LQ_2$ signal,
and for the data.
The number of events in the four signal bins is shown in Fig.\
\ref{fig:lastcut}.

\begin{table*} 
\caption{\label{tab:cuflow} Signal efficiency ($\varepsilon$) for
  various scalar leptoquark masses, 
  number of expected background events
  ($N_{\text{pred}}^{\text{bgd}}$), and the number of
  data events ($N_{\text{data}}$). 
}
\begin{center}
\begin{ruledtabular}
\begin{tabular}{lccccr} 
  Cut & $m(\mu\mu)>105\gev$ & Bin 0 & Bin 1 & Bin 2 & Bin 3 \\ 
  \hline
  $\varepsilon(m_{\LQ_2}=140\gev)$ &$0.139\pm0.013$ &$0.041\pm0.004$ &$0.036\pm0.004$ &$0.025\pm0.003$ &$0.038\pm0.005$ \\
  $\varepsilon(m_{\LQ_2}=160\gev)$ &$0.174\pm0.016$ &$0.026\pm0.004$ &$0.042\pm0.004$ &$0.040\pm0.005$ &$0.067\pm0.008$ \\
  $\varepsilon(m_{\LQ_2}=180\gev)$ &$0.197\pm0.018$ &$0.017\pm0.002$ &$0.038\pm0.004$ &$0.049\pm0.005$ &$0.093\pm0.011$ \\
  $\varepsilon(m_{\LQ_2}=200\gev)$ &$0.215\pm0.019$ &$0.009\pm0.002$ &$0.026\pm0.004$ &$0.047\pm0.005$ &$0.133\pm0.015$ \\
  $\varepsilon(m_{\LQ_2}=220\gev)$ &$0.223\pm0.020$ &$0.005\pm0.001$ &$0.016\pm0.003$ &$0.039\pm0.005$ &$0.163\pm0.017$ \\
  $\varepsilon(m_{\LQ_2}=240\gev)$ &$0.243\pm0.021$ &$0.005\pm0.001$ &$0.013\pm0.002$ &$0.032\pm0.004$ &$0.193\pm0.018$ \\
  $\varepsilon(m_{\LQ_2}=260\gev)$ &$0.251\pm0.022$ &$0.004\pm0.001$ &$0.009\pm0.002$ &$0.025\pm0.004$ &$0.212\pm0.019$ \\
  $\varepsilon(m_{\LQ_2}=280\gev)$ &$0.256\pm0.022$ &$0.003\pm0.001$ &$0.006\pm0.001$ &$0.018\pm0.003$ &$0.229\pm0.020$ \\
  $\varepsilon(m_{\LQ_2}=300\gev)$ &$0.263\pm0.023$ &$0.004\pm0.001$ &$0.004\pm0.001$ &$0.013\pm0.002$ &$0.242\pm0.021$ \\
  \hline
  $N_{\text{pred}}^{\text{bgd}}$ &$6.760\pm1.999$ &$5.140\pm1.565$
  &$0.958\pm0.374$ &$0.388\pm0.144$ &$0.274\pm0.138$\VSPP\\ 
  \hline
  $N_{\text{data}}$ &$6~~~~~~~~~~~~~~~~$ &$2~~~~~~~~~~~~~~~~$ &$2~~~~~~~~~~~~~~~~$ &$2~~~~~~~~~~~~~~~~$ &$0~~~~~~~~~~~~~~~~$ \\
\end{tabular}
\end{ruledtabular}
\end{center}
\end{table*} 

Table \ref{tab:cuflow} summarizes the efficiencies
for various leptoquark masses, as well as the numbers of expected
background events and the distribution of the data in the four
signal bins.
The signal efficiency increases with mass, 
because for larger leptoquark masses, the decay products
have larger momenta yielding events with larger $S_T$.
The dominant uncertainty on the predicted number of background
events is due to MC statistics and varies between $7\%$ and $25\%$
for the four signal bins. 
Other contributions arise from the 
jet-energy calibration uncertainty ($2\%$ -- $12\%$)
and the uncertainty in the shape of the jet transverse energy distribution
($20\%$),  
which has been estimated by a comparison of the \pythia{}
and \alpgen{} simulations.  
The jet multiplicity in DY events generated with \pythia{}, which is a
leading-order generator, was corrected in order to reflect the 
multiplicity distribution observed
in the data around the $Z$ boson. This was
accomplished by comparing exponential fits to the inclusive jet
multiplicity distribution in data and Monte Carlo. 
The fit is dominated by the zero and one jet bins. The remaining 
difference in the two jet bin between $\mu j\mu j$ events in data 
and in the \pythia{} Monte Carlo  
in the vicinity of the $Z$ boson resonance, $60\gev<m(\mu\mu)<105\gev$, 
was taken as the corresponding systematic uncertainty ($16\%$).
In addition, the following sources of systematic uncertainties 
were taken into account: luminosity ($6.5\%$), 
PDF uncertainty
of the DY processes ($3.6\%$), and muon triggering and 
identification ($5\%$). The systematics, added in quadrature, 
are shown in Table \ref{tab:cuflow}.
The systematic uncertainties on the signal efficiencies 
arise from limited
Monte Carlo statistics ($2\%$ -- $17\%$), jet-energy scale
($3\%$ -- $13\%$), muon triggering and identification ($5\%$), and 
PDF uncertainty ($2\%$).

No significant excess of data over background was observed. 
Upper limits on the product of cross section times branching
fraction, $\sigma\cdot\beta^2$, were calculated 
as described in
reference \cite{tomjunk},
by treating the four signal bins as individual channels.
The likelihoods for the different bins
were combined with correlations of systematic
uncertainties 
taken into account.
The
limits are calculated using the confidence level $CL_S =
CL_{S+B}/CL_{B}$, where $CL_{S+B}$ is the confidence level
for the signal plus background hypothesis and $CL_{B}$ is the
confidence level for the background only \cite{tomjunk}. 

\begin{figure}
\includegraphics[width=\picwidth]{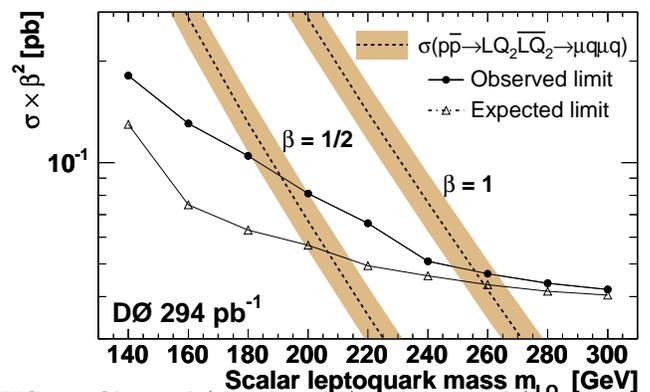}
\vspace*{-0.6cm} 
\caption{\label{fig:limits}
  Observed (closed circles) and expected (open triangles) $95\,\%$ 
  C.L.\ upper limit on production cross section times branching 
  fraction for second generation scalar leptoquarks. The NLO theoretical 
  predictions are also shown with error bands for $\beta = 1$ and $1/2$.
}
\end{figure}

The limits on the cross section times branching fraction and
the theoretical predictions \cite{kraemer1} 
are shown in Fig.\ \ref{fig:limits} and Table \ref{tab:combination},
as well as the average expected limit assuming 
that no signal is present.
Due to the larger background, the contribution of bin 0 to the limit
is relatively small. This explains why the average 
expected limit is better than the
observed limit, although the sum of the events in all four bins is comparable
to the background prediction.
The mass limit is extracted from the intersection of the lower edge of
the  next-to-leading order (NLO) 
cross section uncertainty band with the observed upper bound on the
cross section. The uncertainty band 
reflects
the PDF uncertainty \cite{cteq6}
as well as the variation of the factorization and renormalization scale
between $m_{\LQ_2}/2$ and $2m_{\LQ_2}$, added in quadrature.

\begin{table*}
\caption{ \label{tab:combination}
  NLO cross sections for 
  scalar leptoquark pair production in $p\bar{p}$ collisions 
  at $\sqrt{s}=1.96\,\rm{TeV}$, expected and observed
  $95\%$ C.L.\ upper limits 
  on the cross section times branching fraction
  for the analysis described in this Letter,
  and observed upper limits for the Run I + Run II combination. 
  The cross sections
  shown are calculated using CTEQ6.1M as PDF \cite{cteq6} 
  and $m_{\LQ_2}$  
  as the factorization/renormalization scale \cite{kraemer1}. The
  uncertainties in the theoretical cross sections 
  originate from a variation of the 
  renormalization and factorization scale between $m_{\LQ_2}/2$
  and $2 m_{\LQ_2}$ and the PDF errors, added in quadrature.
}
\begin{ruledtabular}
\begin{tabular}{lcccr}
$m_{\LQ_2}$ & 
$\sigma_{\text{theory}}^{\text{Run II}}~[\rm{pb}]$ &
\multicolumn{2}{c}{Run II limits on $\sigma\cdot\beta^2~[\rm{pb}]~~~~~~~~~~~~$}& 
Run I\,+\,II limits  \\
$[\rm{GeV}]$ & 
$\sqrt{s}=1.96\,\rm{TeV}$ &
(expected) & 
(observed) &
on $\sigma\cdot\beta^2~[\rm{pb}]$
\\
\hline \vspace{-0.3cm} \\
$140$ & 
$2.380_{-0.448}^{+0.487}$ & $0.130$ & $0.181$ & $0.144\SP$
\VSP\\
$160$ & 
$1.080_{-0.200}^{+0.225}$ & $0.075$ & $0.131$ &$0.104\SP$
\VSP\\
$180$ & 
$0.525_{-0.096}^{+0.111}$ & $0.063$ & $0.105$ &$0.083\SP$
\VSP\\
$200$ & 
$0.268_{-0.049}^{+0.057}$ & $0.057$ & $0.081$ &$0.064\SP$
\VSP\\
$220$ & 
$0.141_{-0.025}^{+0.030}$ & $0.049$ & $0.066$ &$0.052\SP$
\VSP\\
$240$ & 
$0.076_{-0.015}^{+0.017}$ & $0.046$ & $0.051$ &$0.045\SP$
\VSP\\
$260$ & 
$0.042_{-0.008}^{+0.009}$ & $0.043$ & $0.047$ &$0.042\SP$
\VSP\\
$280$ & 
$0.023_{-0.004}^{+0.005}$ & $0.042$ & $0.044$ &$0.038\SP$
\VSP\\
$300$ & 
$0.013_{-0.002}^{+0.003}$ & $0.040$ & $0.042$ &$0.037~\SP$
\\
\end{tabular}
\end{ruledtabular}
\end{table*}

\begin{figure}
\includegraphics[width=\picwidth]{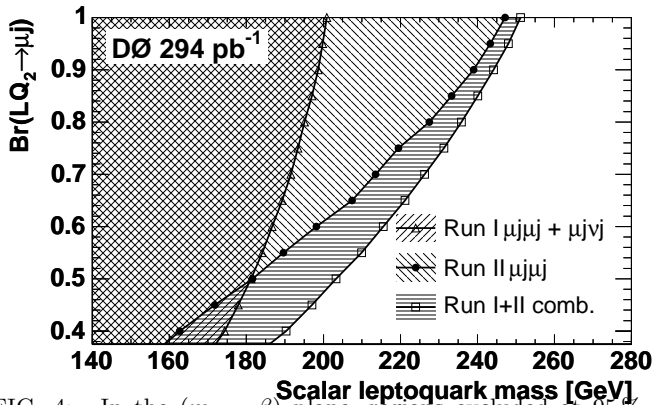}
\vspace*{-0.6cm} 
\caption{\label{fig:exclusionplot} 
  In the ($m_{\LQ_2}$,$\beta$) plane, regions excluded at 
  $95\,\%$ C.L.\ by the D0 Run I results, by this analysis, 
  and by the combination of the two.
}
\end{figure}

The lower limit on the mass of second generation scalar leptoquarks 
was determined at the $95\%$ C.L. to be 
\masslimit{} and \masslimitbetahalf{} for 
$\beta=1$ and $\beta=1/2$, respectively. 
The average expected limits 
are \expectedmasslimit{} and \expectedmasslimithalf{}.
Figure \ref{fig:exclusionplot} shows the excluded region in the
$\beta$ versus $m_{\LQ_2}$ parameter space. 

The D\O\ Run I analysis in the $\mu j\mu j$ channel 
had no events after all cuts, 
while $0.7\pm0.5$ events were expected from the background.
A complementary Run I analysis in the $\mu j\nu j$ channel yielded 
no events for 
$0.7\pm0.9$ events expected from standard model background 
\cite{run1_lq2pairs}. 
Taking into account the smaller cross section for the production of
second generation scalar leptoquarks
at the Run I center-of-mass energy 
$\sqrt{s}=1.8\,\rm{TeV}$, these earlier results 
have been combined with the Run II analysis
presented in this Letter. The results are summarized in 
Table \ref{tab:combination} and 
the excluded parameter regions are shown in 
Fig.\ \ref{fig:exclusionplot}. The combined lower limit for scalar
leptoquarks of the second generation is \combmasslimit{} 
(\combmasslimitbetahalf{}) for $\beta=1$ ($\beta=1/2$).
These results improve on previous measurements at the Tevatron
collider \cite{run1_lq2pairs,run2_cdf}   
and are, for large $\beta$, 
the most stringent limits on second generation scalar
leptoquarks from direct measurements to date.



\input acknowledgement_paragraph_r2.tex   

\end{document}

%% file: list_of_authors_r2.tex
%
\author{                                                                      
\vspace{-0.3cm}
V.M.~Abazov,$^{36}$                                                           
B.~Abbott,$^{75}$                                                             
M.~Abolins,$^{65}$                                                            
B.S.~Acharya,$^{29}$                                                          
M.~Adams,$^{52}$                                                              
T.~Adams,$^{50}$                                                              
M.~Agelou,$^{18}$                                                             
J.-L.~Agram,$^{19}$                                                           
S.H.~Ahn,$^{31}$                                                              
M.~Ahsan,$^{59}$
% ...
\\                                                                            
\vskip 0.30cm                                                                 
\centerline{(D\O\ Collaboration)}                                             
\vskip 0.30cm                                                                 
}                                                                             
\affiliation{                                                                 
\centerline{$^{1}$Universidad de Buenos Aires, Buenos Aires, Argentina}       
\centerline{$^{2}$LAFEX, Centro Brasileiro de Pesquisas F{\'\i}sicas,         
                  Rio de Janeiro, Brazil}                                     
\centerline{$^{3}$Universidade do Estado do Rio de Janeiro,                   
                  Rio de Janeiro, Brazil}                                     
\centerline{$^{4}$Instituto de F\'{\i}sica Te\'orica, Universidade            
                  Estadual Paulista, S\~ao Paulo, Brazil}                     
\centerline{$^{5}$University of Alberta, Edmonton, Alberta, Canada,           
               Simon Fraser University, Burnaby, British Columbia, Canada,}   
\centerline{York University, Toronto, Ontario, Canada, and                    
         McGill University, Montreal, Quebec, Canada}                         
\centerline{$^{6}$Institute of High Energy Physics, Beijing,                  
                  People's Republic of China}                                 
\centerline{$^{7}$University of Science and Technology of China, Hefei,       
                  People's Republic of China}                                 
\centerline{$^{8}$Universidad de los Andes, Bogot\'{a}, Colombia}             
\centerline{$^{9}$Center for Particle Physics, Charles University,            
                  Prague, Czech Republic}                                     
\centerline{$^{10}$Czech Technical University, Prague, Czech Republic}        
\centerline{$^{11}$Center for Particle Physics, Institute of Physics,         
                   Academy of Sciences of the Czech Republic,                 
                   Prague, Czech Republic}                                    
\centerline{$^{12}$Universidad San Francisco de Quito, Quito, Ecuador}        
\centerline{$^{13}$Laboratoire de Physique Corpusculaire, IN2P3-CNRS,         
                  Universit\'e Blaise Pascal, Clermont-Ferrand, France}       
\centerline{$^{14}$Laboratoire de Physique Subatomique et de Cosmologie,      
                  IN2P3-CNRS, Universite de Grenoble 1, Grenoble, France}     
\centerline{$^{15}$CPPM, IN2P3-CNRS, Universit\'e de la M\'editerran\'ee,     
                  Marseille, France}                                          
\centerline{$^{16}$IN2P3-CNRS, Laboratoire de l'Acc\'el\'erateur              
                  Lin\'eaire, Orsay, France}                                  
\centerline{$^{17}$LPNHE, IN2P3-CNRS, Universit\'es Paris VI and VII,         
                  Paris, France}                                              
\centerline{$^{18}$DAPNIA/Service de Physique des Particules, CEA, Saclay,    
                  France}                                                     
\centerline{$^{19}$IReS, IN2P3-CNRS, Universit\'e Louis Pasteur, Strasbourg,  
                France, and Universit\'e de Haute Alsace, Mulhouse, France}   
\centerline{$^{20}$Institut de Physique Nucl\'eaire de Lyon, IN2P3-CNRS,      
                   Universit\'e Claude Bernard, Villeurbanne, France}         
\centerline{$^{21}$III. Physikalisches Institut A, RWTH Aachen,               
                   Aachen, Germany}                                           
\centerline{$^{22}$Physikalisches Institut, Universit{\"a}t Bonn,             
                  Bonn, Germany}                                              
\centerline{$^{23}$Physikalisches Institut, Universit{\"a}t Freiburg,         
                  Freiburg, Germany}                                          
\centerline{$^{24}$Institut f{\"u}r Physik, Universit{\"a}t Mainz,            
                  Mainz, Germany}                                             
\centerline{$^{25}$Ludwig-Maximilians-Universit{\"a}t M{\"u}nchen,            
                   M{\"u}nchen, Germany}                                      
\centerline{$^{26}$Fachbereich Physik, University of Wuppertal,               
                   Wuppertal, Germany}                                        
\centerline{$^{27}$Panjab University, Chandigarh, India}                      
\centerline{$^{28}$Delhi University, Delhi, India}                            
\centerline{$^{29}$Tata Institute of Fundamental Research, Mumbai, India}     
\centerline{$^{30}$University College Dublin, Dublin, Ireland}                
\centerline{$^{31}$Korea Detector Laboratory, Korea University,               
                   Seoul, Korea}                                              
\centerline{$^{32}$SungKyunKwan University, Suwon, Korea}                     
\centerline{$^{33}$CINVESTAV, Mexico City, Mexico}                            
\centerline{$^{34}$FOM-Institute NIKHEF and University of                     
                  Amsterdam/NIKHEF, Amsterdam, The Netherlands}               
\centerline{$^{35}$Radboud University Nijmegen/NIKHEF, Nijmegen, The          
                  Netherlands}                                                
\centerline{$^{36}$Joint Institute for Nuclear Research, Dubna, Russia}       
\centerline{$^{37}$Institute for Theoretical and Experimental Physics,        
                  Moscow, Russia}                                             
\centerline{$^{38}$Moscow State University, Moscow, Russia}                   
\centerline{$^{39}$Institute for High Energy Physics, Protvino, Russia}       
\centerline{$^{40}$Petersburg Nuclear Physics Institute,                      
                   St. Petersburg, Russia}                                    
\centerline{$^{41}$Lund University, Lund, Sweden, Royal Institute of          
                   Technology and Stockholm University, Stockholm,            
                   Sweden, and}                                               
\centerline{Uppsala University, Uppsala, Sweden}                              
\centerline{$^{42}$Physik Institut der Universit{\"a}t Z{\"u}rich,            
                    Z{\"u}rich, Switzerland}                                  
\centerline{$^{43}$Lancaster University, Lancaster, United Kingdom}           
\centerline{$^{44}$Imperial College, London, United Kingdom}                  
\centerline{$^{45}$University of Manchester, Manchester, United Kingdom}      
\centerline{$^{46}$University of Arizona, Tucson, Arizona 85721, USA}         
\centerline{$^{47}$Lawrence Berkeley National Laboratory and University of    
                  California, Berkeley, California 94720, USA}                
\centerline{$^{48}$California State University, Fresno, California 93740, USA}
\centerline{$^{49}$University of California, Riverside, California 92521, USA}
\centerline{$^{50}$Florida State University, Tallahassee, Florida 32306, USA} 
\centerline{$^{51}$Fermi National Accelerator Laboratory, Batavia,            
                   Illinois 60510, USA}                                       
\centerline{$^{52}$University of Illinois at Chicago, Chicago,                
                   Illinois 60607, USA}                                       
\centerline{$^{53}$Northern Illinois University, DeKalb, Illinois 60115, USA} 
\centerline{$^{54}$Northwestern University, Evanston, Illinois 60208, USA}    
\centerline{$^{55}$Indiana University, Bloomington, Indiana 47405, USA}       
\centerline{$^{56}$University of Notre Dame, Notre Dame, Indiana 46556, USA}  
\centerline{$^{57}$Iowa State University, Ames, Iowa 50011, USA}              
\centerline{$^{58}$University of Kansas, Lawrence, Kansas 66045, USA}         
\centerline{$^{59}$Kansas State University, Manhattan, Kansas 66506, USA}     
\centerline{$^{60}$Louisiana Tech University, Ruston, Louisiana 71272, USA}   
\centerline{$^{61}$University of Maryland, College Park, Maryland 20742, USA} 
\centerline{$^{62}$Boston University, Boston, Massachusetts 02215, USA}       
\centerline{$^{63}$Northeastern University, Boston, Massachusetts 02115, USA} 
\centerline{$^{64}$University of Michigan, Ann Arbor, Michigan 48109, USA}    
\centerline{$^{65}$Michigan State University, East Lansing, Michigan 48824,   
                   USA}                                                       
\centerline{$^{66}$University of Mississippi, University, Mississippi 38677,  
                   USA}                                                       
\centerline{$^{67}$University of Nebraska, Lincoln, Nebraska 68588, USA}      
\centerline{$^{68}$Princeton University, Princeton, New Jersey 08544, USA}    
\centerline{$^{69}$State University of New York, Buffalo, New York 14260, USA}
\centerline{$^{70}$Columbia University, New York, New York 10027, USA}        
\centerline{$^{71}$University of Rochester, Rochester, New York 14627, USA}   
\centerline{$^{72}$State University of New York, Stony Brook,                 
                   New York 11794, USA}                                       
\centerline{$^{73}$Brookhaven National Laboratory, Upton, New York 11973, USA}
\centerline{$^{74}$Langston University, Langston, Oklahoma 73050, USA}        
\centerline{$^{75}$University of Oklahoma, Norman, Oklahoma 73019, USA}       
\centerline{$^{76}$Oklahoma State University, Stillwater, Oklahoma 74078, USA}
\centerline{$^{77}$Brown University, Providence, Rhode Island 02912, USA}     
\centerline{$^{78}$University of Texas, Arlington, Texas 76019, USA}          
\centerline{$^{79}$Southern Methodist University, Dallas, Texas 75275, USA}   
\centerline{$^{80}$Rice University, Houston, Texas 77005, USA}                
\centerline{$^{81}$University of Virginia, Charlottesville, Virginia 22901,   
                   USA}                                                       
\centerline{$^{82}$University of Washington, Seattle, Washington 98195, USA}  
\vspace{-0.3cm}
}                                                                             

%% file: acknowledgement_paragraph_r2.tex
%
We thank the staffs at Fermilab and collaborating institutions, 
and acknowledge support from the 
DOE and NSF (USA);
CEA and CNRS/IN2P3 (France);
FASI, Rosatom and RFBR (Russia);
CAPES, CNPq, FAPERJ, FAPESP and FUNDUNESP (Brazil);
DAE and DST (India);
Colciencias (Colombia);
CONACyT (Mexico);
KRF and KOSEF (Korea);
CONICET and UBACyT (Argentina);
FOM (The Netherlands);
PPARC (United Kingdom);
MSMT (Czech Republic);
CRC Program, CFI, NSERC and WestGrid Project (Canada);
BMBF and DFG (Germany);
SFI (Ireland);
The Swedish Research Council (Sweden);
Research Corporation;
Alexander von Humboldt Foundation;
and the Marie Curie Program.